# Article title

Open STM: A low-cost scanning tunneling microscope with a fast approach method

# Authors

Weilin Ma

# Affiliations

University of Chinese Academy of Sciences, Beijing, 100049, China

Institute of Optics and Electronics, Chinese Academy of Sciences, Chengdu, 610209, China

# Corresponding author's email address and Twitter handle

maweilin22@mails.ucas.edu.cn

# Abstract

In this paper, we have designed a low-cost scanning tunneling microscope (STM) priced at 300 USD or 2000 CNY. This microscope is suitable for educational purposes and low-demand research imaging at the nanometer level. This microscope's motion components and scanner are controlled using piezoelectric materials, avoiding the thermal drift associated with traditional motor control. Our tip approach algorithm, which considers the capacitance and friction characteristics during piezoelectric slider movement, has reduced the time required for sample loading to establish tunneling current to approximately 1 minute. The dimensions of the microscope body are 45×45×31.5mm(W×L×H), and the control voltage does not exceed 15V, ensuring the safety of operators with limited experience. In the performance verification, we performed a scanning tunneling scan on a Highly Oriented Pyrolytic Graphite(HOPG) sample with bias voltages of 50mV and 60mV, resulting in clear observations of the atomic features of HOPG in the STM pattern.

**Specifications table**

| Hardware name | Open STM(Open-sources scanning tunneling microscope) |
|---|---|
| Subject area | - Engineering and materials science<br>- Educational tools and open source alternatives to existing infrastructure |
| Hardware type | Imaging tools |
| Closest commercial analog | No commercial analog is available. |
| Open source license | CC-BY-SA-4.0 and GPLv3(Software, / Python files) License |
| Cost of hardware | 300USD or 2000CNY |
| Source file repository | https://doi.org/10.17632/f35c6xzzcm.1 |

# 1 Hardware in context

Since 1981, when Binnig and Rohrer from IBM invented the Scanning Tunneling Microscope(STM) and was awarded the Nobel Prize in 1986[1][2], the STM has found widespread applications in the field of science for its high performance in sub-nanometer scale imaging. The STM can visualize individual atoms and molecules on various materials, offering insights into their arrangement, organization, and defects. However, such an instrument is inaccessible to the general public for its technical expertise, and it is usually purchased by universities and research institutions through scientific instruments companies for $8,000 (low cost) to $30,000 -150,000(professional)[3].

In 1986, Besocke proposed an easily operable STM design[4], which does not require a vacuum and low-temperature environment. It utilizes a tripod structure built with piezoelectric tubes to achieve STM imaging. Although the piezoelectric tube used in it is still relatively expensive up to the present time, the simplicity and stability design principles proposed in the paper can serve as core principles for subsequent designs. Alexander developed a simple STM design on his website in about 2001[5], which makes DIY a low-cost STM possible. He invented a new type of scanner called the "unimorph disk scanner" based on his patent[6], which involves cutting a piezoelectric buzzer, costing around 1$, into four parts and applying a voltage to these sections individually to achieve three-dimensional control. The introduction of this type of scanner significantly reduced costs. In 2015, Berard published his home-built STM[7] with an atomic resolution, and based on the design file, we can reproduce the STM with fair cost and difficulties. Berard used Alexander's scanner design and employed a tripod lifting structure with a stepper motor to accomplish the coarse approach procedure. However, since this structure is open-loop, it cannot estimate the distance between the sample and the tip. To avoid tip crashing, such approach procedures typically require considerable time. Additionally, the heating from the stepper motor leads to thermal expansion of the microscope structure, causing instability in the distance between the sample and the tip. Liao et al.[8] unveiled a low-cost nanopositioner that utilizes the stick-slip effect to achieve nanometer-level movement in 2022. Moreover, the positioning slider does not generate heat due to using piezoelectric ceramics as actuators. Using this positioner as a coarse approach mechanism for STM can improve the issues of approach speed and thermal drift.

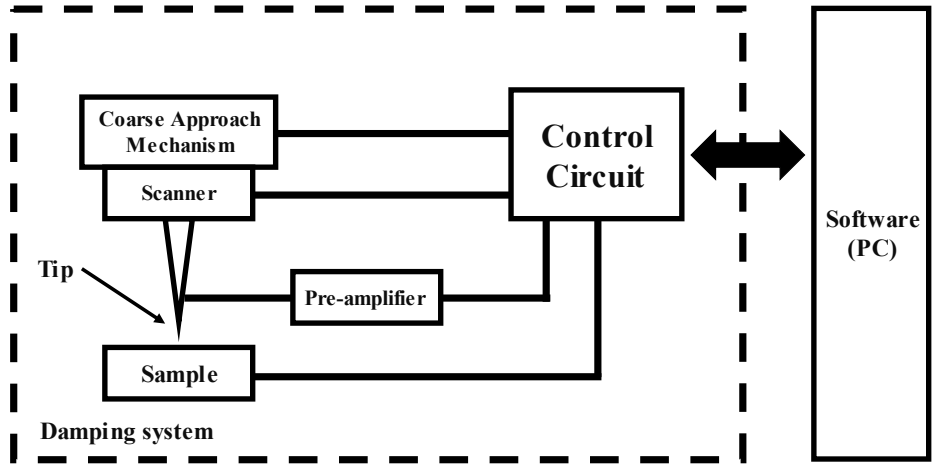

Fig. 1. Schematic of a STM

During the imaging process in STM, a bias voltage is applied to the sample. The sample and the tip used are brought from each other into a tunneling distance (typically 4-7 Å[9]). A piezoelectric component is used to control the tip for scanning and imaging. At this point, the tunneling current flowing between the sample and the tip can reflect changes in the surface morphology of the sample. To achieve this process, the designation of STM mainly includes these parts(As shown in Fig. 1):

- **Scanner:** The scanner enables controlled movement in three axes(XYZ) of the tip, facilitating translational scanning of the sample surface.
- **Coarse approach mechanism:** The coarse approach mechanism is responsible for controlling the movement of the sample towards the tip, ensuring that the sample is positioned within the scanning range of the scanner.
- **Vibration isolation system:** Maintaining a stable distance between the tip and the sample during scanning is crucial for accurate measurements. Two effective designs are commonly employed for STM systems: spring suspension vibration isolation or a vibration isolation platform composed of Viton and stacked metal plate[10][11][12]. Magnetic damping can further enhance vibration isolation in the suspension system.
- **Pre-amplifier:** Tunneling currents typically range from picoamps to nanoamps[13]. A pre-amplifier is necessary to detect such minuscule currents using typical circuits. Common designs include Transimpedance Amplifiers (TIA) and instrumentation amplifiers[14].
- **Control circuit:** The control circuit controls the various components of the STM system and facilitates communication with a computer. This includes the transmission of commands, signals, and other pertinent information.
- **Software:** Operational software is indispensable for controlling the STM and facilitating image processing and storage. This software plays a vital role in operating and utilizing the microscope system.

Building upon the work of Alexander, Berard, and Liao et al., we have developed a low-cost STM(300USD or 2000CNY). The microscope utilizes complete piezoelectric control, effectively mitigating thermal drift issues. Additionally, we have identified an electrical signal during the coarse approach process that serves as a reference for the tip-sample distance, significantly reducing the time required to establish tunneling current from sample loading(typically around 1 minute).

## 2 Hardware description

This paper presents an easy-to-operate, cost-effective STM design capable of achieving atomic-level imaging of Highly Oriented Pyrolytic Graphite (HOPG) surfaces. The microscope comprises a control unit, an STM body, and a vibration-damping system, as shown in Fig. 2(a). The dimensions of the microscope body are 65×65×55mm(W×L×H), and the core unit, excluding the dust cover, measures 45×45×31.5mm(W×L×H, shown in Fig. 2(b)). Motion control of the STM is achieved using piezoelectric materials in conjunction with our control algorithms. This significantly reduces the time required for a coarse approach to just 1 minute, eliminating the need for additional external optical microscopes and capacitance detection circuits[15]. Furthermore, the system operates at a maximum voltage of not exceeding 15V, virtually eliminating the risk of electrical shock during operation and providing safety for less experienced operators. Our software(Fig. 3) offers basic imaging functionality and includes various testing functions, including tunneling distance-current curve testing, bias voltage testing, scanning head repeatability testing, and more. Test results are displayed and saved in real-time.

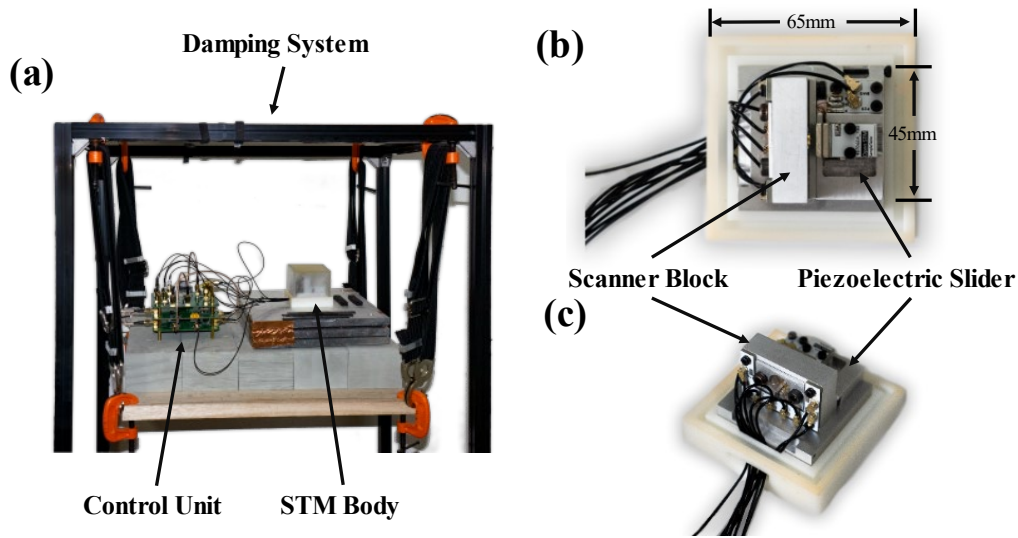

Fig. 2. The assembled STM. (a) The microscope body and control unit are placed on a simple damping system. (b) Top view of the STM body with the dust cover open. (c) Side view of the STM body.

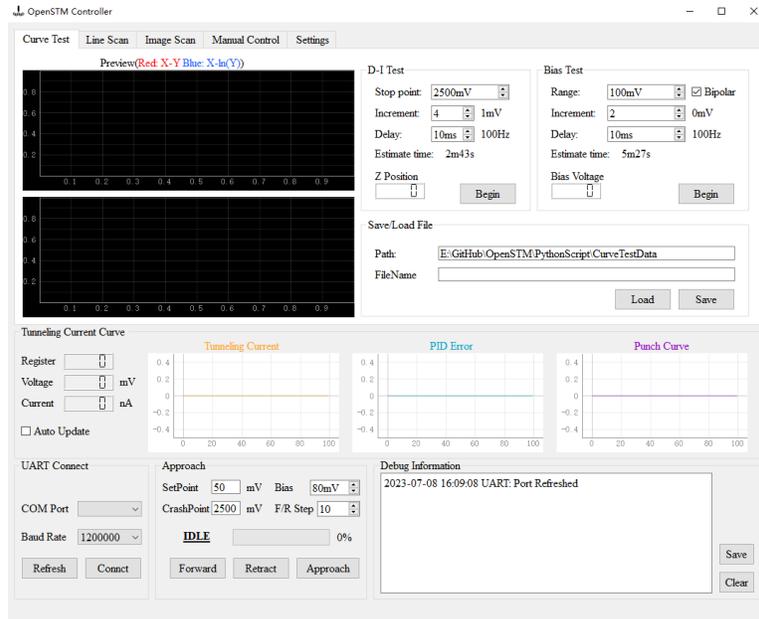

Fig. 3. PC software interface

Our STM can be applied in the field of education as well as in research areas(low demand).

- In education, this device can demonstrate the tunneling principle in quantum mechanics to students. Additionally, since reproducing this project is relatively easy, it can be incorporated as a part of the teaching process, allowing students to gain practical experience in piezoelectric control, soldering skills, and equipment debugging.

- In the research field, this equipment can provide high-magnification qualitative imaging of sample surfaces and the capability to perform tunneling bias voltage curve tests.

## 3   Design files summary

| Design file name | File type | Open source license | Location of the file |
|---|---|---|---|
| 3DModels | STEP, PDF, SLDPRT, SLDASM files | CC-BY-SA-4.0 | https://doi.org/10.17632/f35c6xzzcm.1 |
| HardwareCode | BIN, ELF, and VS code project files | CC-BY-SA-4.0 | https://doi.org/10.17632/f35c6xzzcm.1 |
| PCB | ZIP, PDF, and Easy EDA project files | CC-BY-SA-4.0 | https://doi.org/10.17632/f35c6xzzcm.1 |
| PythonScript | Python files | GPLv3 | https://doi.org/10.17632/f35c6xzzcm.1 |

- 3D Models

The STM body is predominantly manufactured through CNC machining of aluminum blocks. A total of 7 components need to be machined. We provide the original Solidworks design and STEP files for manufacturing in this folder. Drill information is documented in the corresponding PDF files.

- PCB

The STM includes eight PCBs, three utilized for the microscope's control unit, one for the pre-amplifier, and four for circuitry interconnection. The PCBs were designed using EasyEDA(Or JLC EDA in China), and this folder includes project files, schematics, and Gerber files for PCB manufacturing.

- HardwareCode

The central controller for the circuit control section employs the ESP32-WROOM-32E-N8 microcontroller module. This folder contains the microcontroller firmware and source code (written using the ESP-IDF framework with PlatformIO).

- PythonScript

This folder includes supervisory software written in Python. Install the libraries(with specific versions) and launch the software by running "launch.bat" or via the command line "python main.py".

## 4 Bill of materials summary

| Designator | Component | Number | Cost per unit - CNY | Total cost - CNY | Source of materials | Material type |
|---|---|---|---|---|---|---|
| Control Board | See PCB/BOM_ControlBoard.xlsx | 1 | 787.46 | 787.46 | LCSC and Mouser | Semiconductor |
| MCU Board | See PCB/BOM_MCUBoard.xlsx | 1 | 32.28 | 32.28 | LCSC | Semiconductor |
| Power Board | See PCB/BOM_PowerBoard.xlsx | 1 | 110.82 | 110.82 | LCSC | Semiconductor |
| Pre-amplifier | See PCB/BOM_PreAmp.xlsx | 1 | 237.61 | 237.61 | LCSC and Mouser | Semiconductor |
| Piezo stack | AL1.65 × 1.65 × 5D-4F | 1 | 32.26 | 32.26 | Taobao | Ceramic |
| Linear slider | BSP715 | 1 | 204 | 204 | Taobao | Metal |
| Base | STM CNC Block part1 | 1 | 48.84 | 48.84 | Sanweihou | Metal |
| Preamp_cover | STM CNC Block part2 | 1 | 35.82 | 35.82 | Sanweihou | Metal |
| Scanner_cover | STM CNC Block | 1 | 40.84 | 40.84 | Sanweihou | Metal |

| | part3 | | | | | |
|---|---|---|---|---|---|---|
| Sample_Table | STM CNC Block part4 | 1 | 25.80 | 25.80 | Sanweihou | Metal |
| PZM_Base | STM CNC Block part5 | 1 | 54.01 | 54.01 | Sanweihou | Metal |
| Scanner_Mount | STM CNC Block part6 | 1 | 43.51 | 43.51 | Sanweihou | Metal |
| Shell | STM 3D print component part1 | 1 | 17.23 | 17.23 | Sanweihou | Polymer |
| Shell_cover | STM 3D print component part1 | 1 | 11.83 | 11.83 | Sanweihou | Polymer |
| Magnent(square) | E-ACTD-W10-H2-T2 | 2 | 1.31 | 2.62 | JLCFA | Metal |
| Magnent(round) | ACTA-B2-D3-L2 | 1 | 1.28 | 12.8 | JLCFA | Metal |
| Piezo Buzzer | 7BB-12-9 | 1 | 2.19 | 2.19 | Mouser | Ceramic |
| Pt Wire | Tip(D=0.3cm, L=1cm) | 1 | 7.2 | 7.2 | ARITER | Mental |
| PCBs | Control board, MCU board, power board, pre-amplifier board, and boards used for interfacing | 8 | 20 | 160 | JLCPCB | Inorganic |
| MMCX Connector | KH-MMCX-KE-STM | 1 | 3.52 | 3.52 | LCSC | Metal |
| IPEX Connector | U.FL-R-SMT-1(10) | 9 | 1 | 9 | LCSC | Metal |
| FPC Connector | THD0510-05CL-GF | 1 | 0.5 | 0.5 | LCSC | Polymer |
| FPC Cable | JS05A-05P-030-3-4 | 1 | 0.5 | 0.5 | LCSC | Polymer |
| Coaxial cable A | SMA to IPEX(40cm) | 6 | 14.8 | 88.8 | XINQY | Polymer |
| Coaxial cable B | DOSIN-811-0211A(SMA to SMA) | 3 | 18.08 | 52.24 | LCSC | Polymer |

| IDC Cable | 2×7Pin, 10cm | 1 | 1.05 | 1.05 | [Taobao](Taobao) | Polymer |
|---|---|---|---|---|---|---|
| M3×6 Screw | EDLA-J2-M3-L6 | 6 | 0.11 | 0.66 | [JLCFA](JLCFA) | Metal |
| M2×4 Screw | EDLA-J2-M2-L4 | 7 | 0.16 | 1.12 | [JLCFA](JLCFA) | Metal |
| M2×3 Screw | EDLA-J2-M2-L3 | 2 | 0.26 | 0.52 | [JLCFA](JLCFA) | Metal |
| M3×10 Screw | EDLA-J2-M3-L10 | 4 | 0.1 | 0.4 | [JLCFA](JLCFA) | Metal |
| M5×6 Screw | EDLA-J2-M5-L6 | 4 | 0.12 | 0.48 | [JLCFA](JLCFA) | Metal |
| Hexagon Standoffs | EJLC-M3-L16 | 12 | 0.5 | 6 | [JLCFA](JLCFA) | Metal |

## 5  Build instructions

### 5.1  Damping system set-up

For constructing the damping system, methods including but not limited to suspension and stacked metal plates with Viton can be employed. This article does not provide a specific method, but we offer our construction approach for reference if needed. We used a rectangular metal frame built from aluminum profiles and suspended a 10kg stone using four springs (k=408N/m, g=9.8m/s² for each). Additionally, on the rock, we placed a stacking arrangement consisting of three aluminum plates (20×20×1cm) overlapping with fluoro rubber rings to enhance the damping effect further.

### 5.2  Control circuit construction

In this step, we will utilize the files from the PCB folder.

#### 5.2.1  PCB fabrication

Firstly, we need to accomplish the fabrication of the PCBs by supplying all the PCB manufacturing files from the PCB/Gerber folder to the PCB manufacturer. The table below presents the process parameters employed for fabricating these eight PCBs.

Table. 1. Parameters used in PCB manufacturing.

| PCB file name | Material | Board Thickness(mm) |
|---|---|---|
| Gerber_ControlBoard | FR-4 | 1.6 |
| Gerber_MCUBoard | FR-4 | 1.6 |
| Gerber_PowerBoard | FR-4 | 1.6 |
| Gerber_PreAmp | FR-4 | 1 |
| Gerber_Sample_ConnectBoard | FR-4 | 1.6 |
| Gerber_SampleTable | FR-4 | 1.6 |
| Gerber_ScannerConnector | FR-4 | 1.6 |
| Gerber_ScannerMount | FR-4 | 1.6 |

*5.2.2  PCB assembly*

Please refer to their respective corresponding BOM tables for the soldering of the **Control Board**, **MCU Board**, **PreAmp**, and **Power Board**. The **Control Board**, **MCU Board**, and **Power Board** form the control unit of the STM. They must be secured sequentially using hexagonal standoffs, as shown in Fig. 4. The remaining PCBs used for interfacing purposes should be soldered, as shown in Fig. 5. The remaining PCBs should be mounted on the STM body. The following sections will explain installation methods and the electrical connections between the PCBs.

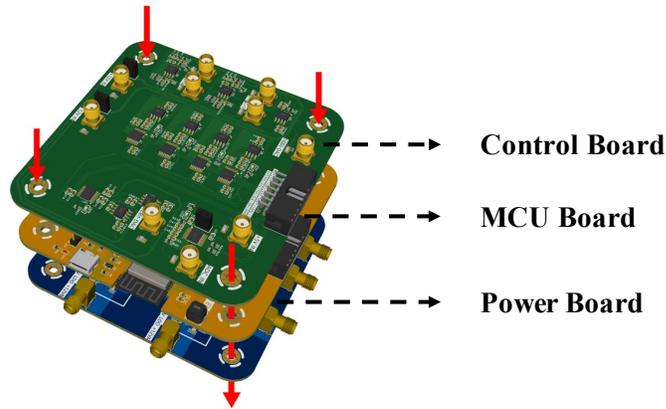

Fig. 4. Installation diagram for the control unit. Secure them using hexagonal standoffs as indicated by the red arrows, paying attention to the stacking order.

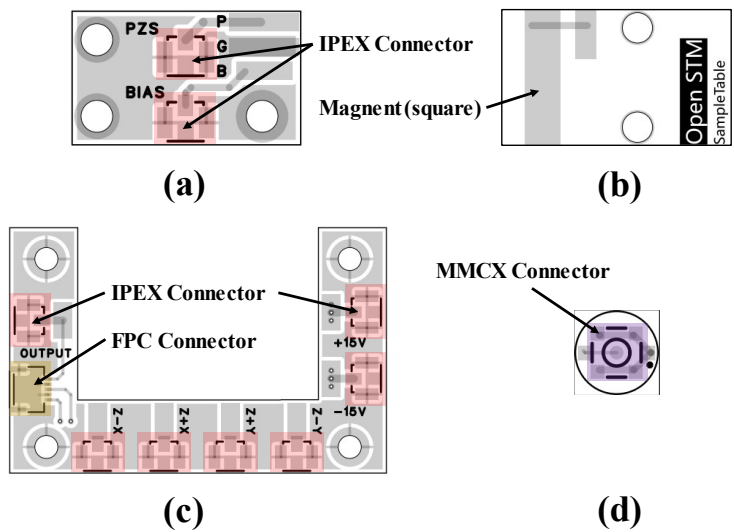

Fig. 5. Soldering Diagram. **(a)** Sample_ConnectBoard, the IPEX connectors in figures (a) and (c) are of the same model. **(b)** SampleTable, when soldering the square magnet, consider using low-temperature solder **(c)** ScannerConnector **(d)** ScannerMount

*5.2.3 Control circuit diagram*

In the STM, the resolution of digital-to-analog and analog-to-digital converters (DAC and ADC) is typically 12 bits or higher[16][17][18]. After considering the trade-off between resolution and cost, we used four 16-bit DACs (AD5761) to control the scanner and apply bias to the sample, one 16-bit ADC (ADS8685) to detect the output signal from the pre-amplifier and one 12-bit DAC (AD5721) to control the piezoelectric slider, as shown in Fig. 6. For amplifying the tunneling current, we employed a simple trans-impedance amplifier(TIA) using an operational amplifier, OPA627, which has a typical input bias current of 1 pA[19]. It can detect and amplify the tunneling current with a feedback resistor of 100 MΩ.

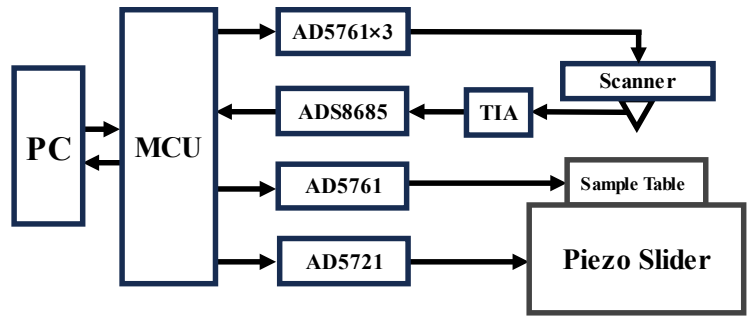

Fig. 6. Block diagram of the control system

## 5.3 STM body assemble

The STM body comprises four parts: scanner block, sample table, piezoelectric slider, and base block. In this step, you must prepare all the STM CNC Blocks(parts 1 to 6). The machining files are located in the 3DModels/CNC folder. Please use 6061 aluminum alloy (or other metal material) for machining, ensuring tolerances are kept within ±0.1mm. Follow the corresponding PDF drawings for tapping as well.

*5.3.1 Scanner*

In this step, we must prepare the piezoelectric buzzer(7BB-12-9), **ScannerMoun**(PCB), and an MMCX connector, as shown in Fig. 7(a). The piezoelectric buzzer typically consists of three layers: a silver-plated layer, a piezoelectric ceramic layer, and a metal layer. (1) To enable three-dimensional movement of the scanner, we need to carefully divide the top silver-plated layer of the buzzer into four quadrants with a cutter knife or any other tools, as shown in Fig. 7(b). (2) Next, we need to solder the MMCX connector onto the **ScannerMount** PCB. (3) After soldering the connector, we also need to solder enamel-coated wires with a diameter of 0.1mm and a length of about 5cm to two metal pads next to the connector. One wire connects to the MMCX connector shell (marked with a dot on the PCB), and the other connects to the MMCX signal end. (4) Use adhesive glue(Egro 5400 in our case) to attach the PCB and the buzzer, ensuring they are concentrically aligned. (5) Finally, use low-temperature solder and enamel-coated wires to solder electrodes onto the four regions of the buzzer.

During the scanner's operation, the buzzer's metal layer will be set to 0V (ground). By applying equal positive or negative voltages to the four silver-plated layers (+X/-X/+Y/-Y, as shown in Fig. 7(b)), the scanner will achieve movement along the Z-axis(shown in Fig. 7(c)). When viewing from the side along the direction indicated by the red line in Fig. 7 (b), observe the effects of applying different voltages to +X/-X: using positive voltage to +X and negative voltage to -X will cause the scanning head to move to the right, and vice versa for leftward movement. The same principle applies to +Y/-Y. By adjusting the voltage across the +X/-X/+Y/-Y quadrants, three-dimensional movement of the scanner can be achieved.

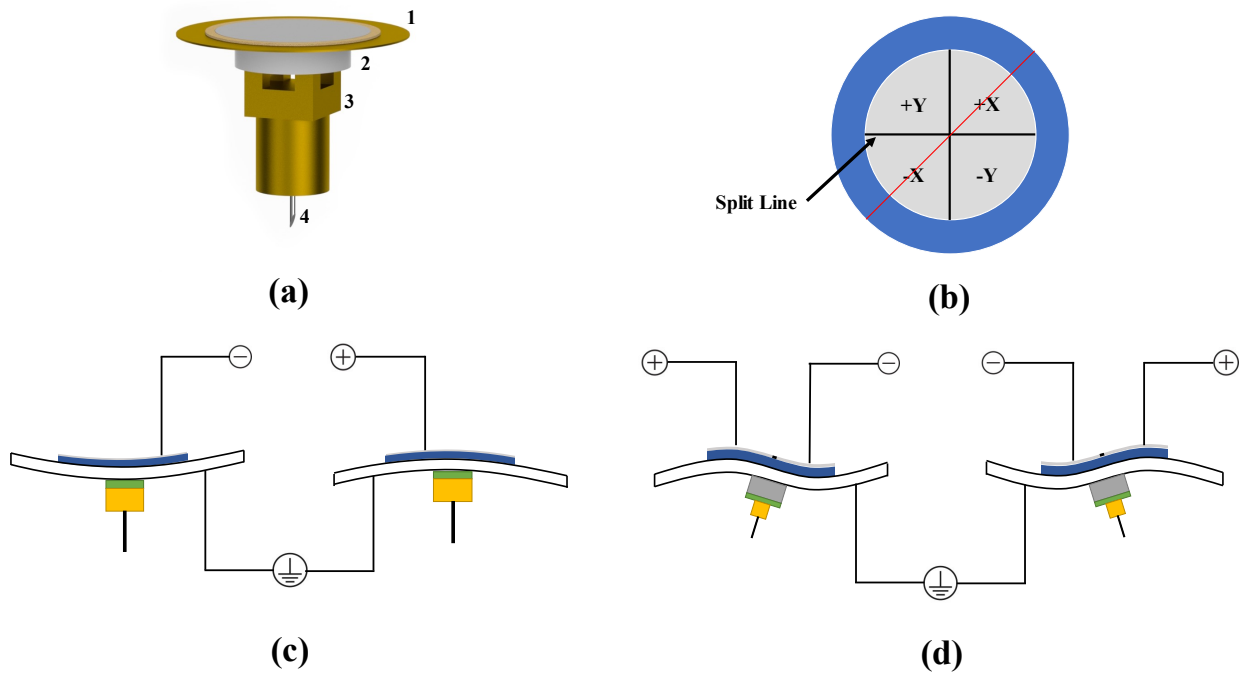

Fig. 7. Composition and operation principle of the scanner. **(a)** Composition of different parts of the scanner.1: Piezoelectric buzzer. 2:ScannerMount PCB. 3: MMCX connector. 4:Tip(Pt wire) **(b)** Illustration of buzzer segmenting. **(c)** Movement of the scanner along the Z-Axis. **(d)** Movement of the scanner along the X/Y-Axis.

### 5.3.2 Scanner Block

To assemble the scanner block, we need to prepare: **ScannerConnector**(PCB), **Preamp_cover** (STM CNC Block part2), **Scanner_cover** (STM CNC Block part3), **Scanner_Mount** (STM CNC Block part6), Scanner (already prepared in the previous step), Pre-amplifier, M3×6 screws(8pcs), and FPC cables.

Assemble the front half of the block as follows: (1)As shown in Fig. 8(b), place the **Scanner_cover** over the scanner and install them into the block, securing it with screws. (2)Use screws to attach the **ScannerConnector** PCB to the block. (3) Solder the electrodes of the scanner onto the corresponding pads on the **ScannerConnector** PCB. Please refer to Fig. 7(b) and the silkscreen markings on the PCB for the correct sequence.

For the rear half of the block assembly, use the **Preamp_cover** to mount the pre-amplifier(**PreAmp**) inside the block and secure it with screws. However, please note that before covering the pre-amplifier with the **Preamp_cover**, you need to insert the FPC cable into the pre-amplifier and connect the other end of the cable, as shown in Fig. 9(a), to the **ScannerConnector** PCB. Additionally, solder the enamel-coated wires from the scanner to their respective positions on the pre-amplifier, as shown in Fig. 9(b). The MMCX connector's shell should be connected to the ground for electromagnetic shielding, and the signal end should be connected to "Current in" for conducting tunneling current.

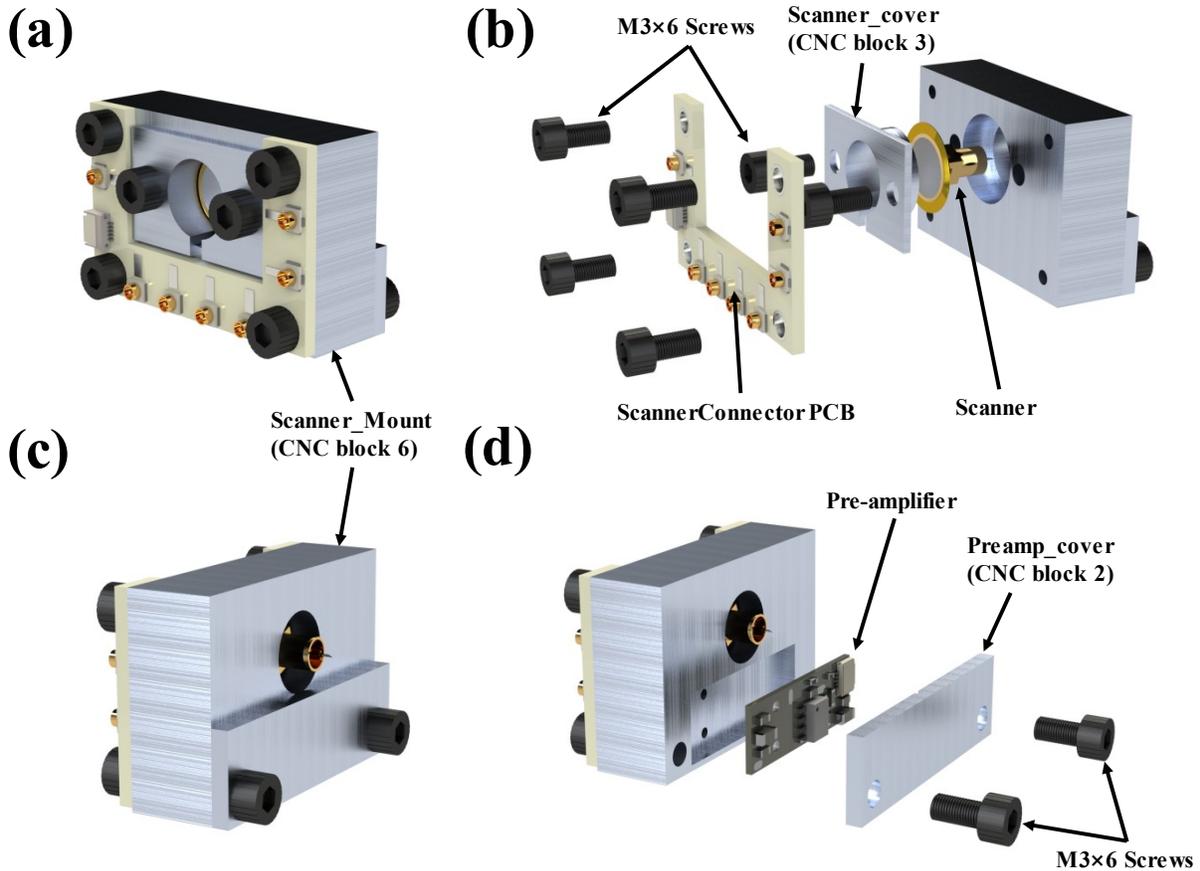

Fig. 8. Assembly diagram of the Scanner Block. (a) Front view of the Scanner Block. (b) Exploded front view of the Scanner Block, showing the need to secure the ScannerConnector PCB and scanner. (c) Rear view of the Scanner Block. (d) Exploded rear view of the Scanner Block, indicating the need to secure the Pre-amplifier in the Scanner_Mount.

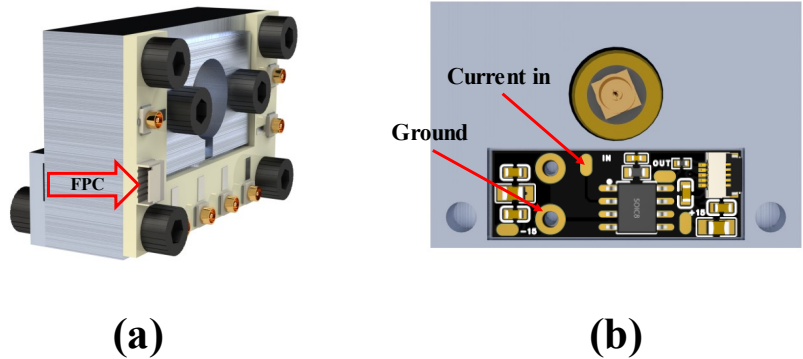

Fig. 9. Wires to be installed on the pre-amplifier. (a) Connected to the ScannerConnector PCB through the FPC cable. (b) The enamel-coated wires from the scanner need to be soldered onto the two indicated solder pads.

*5.3.3  Piezoelectric slider and sample table*

Before assembly, you need to prepare the following components: **Sample Table** (PCB), **Sample_Table** (CNC Block part4), BSP 715 linear slider, magnet (round), magnet (square), piezo stack (AL1.65×1.65×5D-4F), **Sample_ConnectBoard** (PCB), **PZM_Base** (CNC Block part 5), M2×4 screws (7 pcs), M2×3 screws (2 pcs).

Next, proceed with the assembly(Follow as Fig. 10):

(1) Secure the **Sample_ConnectBoard**(PCB) on the **PZM_Base**(CNC Block) using screws.

(2) Use screws to fasten the BSP715 linear slider onto the **PZM_Base**(CNC Block). During this step, ensure the slider is parallel to the installation groove.

(3) Attach the piezo stack to the edge of the **PZM_Base**(CNC Block) using adhesive, as shown in Fig. 10(a) and Fig. 10(b). Then, solder the two electrodes of the piezo stack to the **Sample_ConnectBoard**(PCB), with the red wire connected to G(G means Ground) and the orange wire connected to P(P means positive. However, the voltage output from here will be bipolar).

(4) Affix the magnet(round) to the slider (note: attach the magnet on the side, not on the bottom or top). Push the slider to make contact between the magnet and the piezo stack. Then, use adhesive to secure the magnet to the piezo stack. At this point, the piezo slider assembly is complete.

(5) Secure the **Sample_Table**(CNC Block) onto the slider using screws. Please note that due to tolerances, the screws may be too long. If this happens, you can apply adhesive to the screw heads, acting as spacers once it solidifies.

(6) Fix the **SampleTable**(PCB) onto the **Sample_Table**(CNC), and use enamel-coated wire to connect the smaller square solder pad on the **SampleTable**(CNC) to pad B(B means Bias) on the **Sample_ConnectBoard**(PCB).

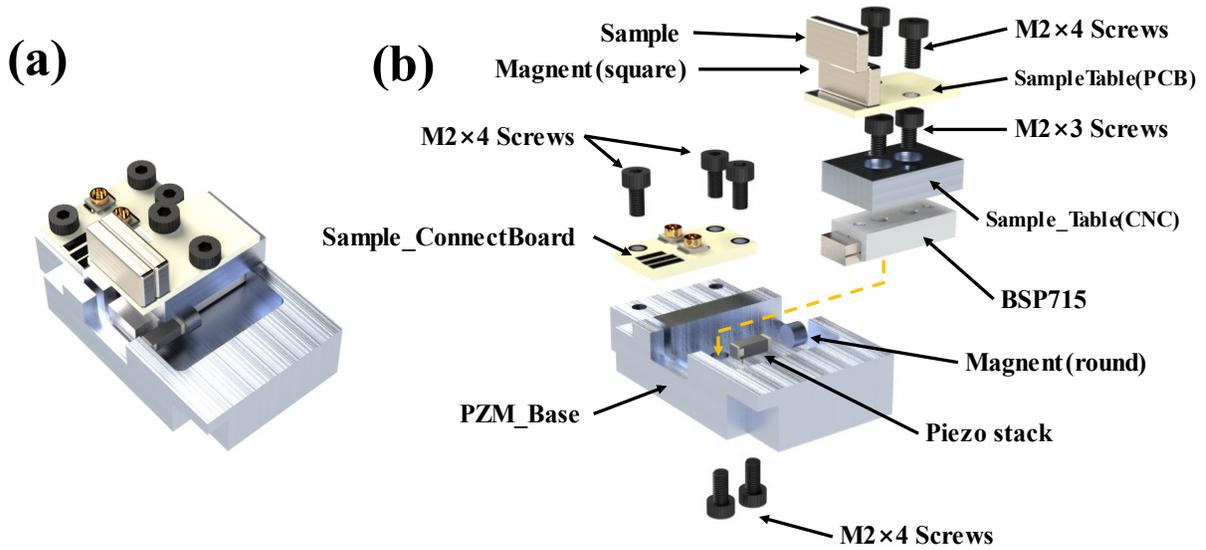

Fig. 10. Assembly diagram of the piezo slider and sample table. (a) Side view (b) Exploded view.

### 5.3.4 Assemble

Complete the final assembly following the sequence as shown in Fig. 11. It's worth noting that when using the M5 screws to secure the piezoelectric slider, they should not be overtightened, as this could potentially alter the magnetic force on the BSP715. The dust cover is not mandatory; removing it can further reduce the microscope's size, but the dust cover effectively prevents dust intrusion and airflow disturbances.

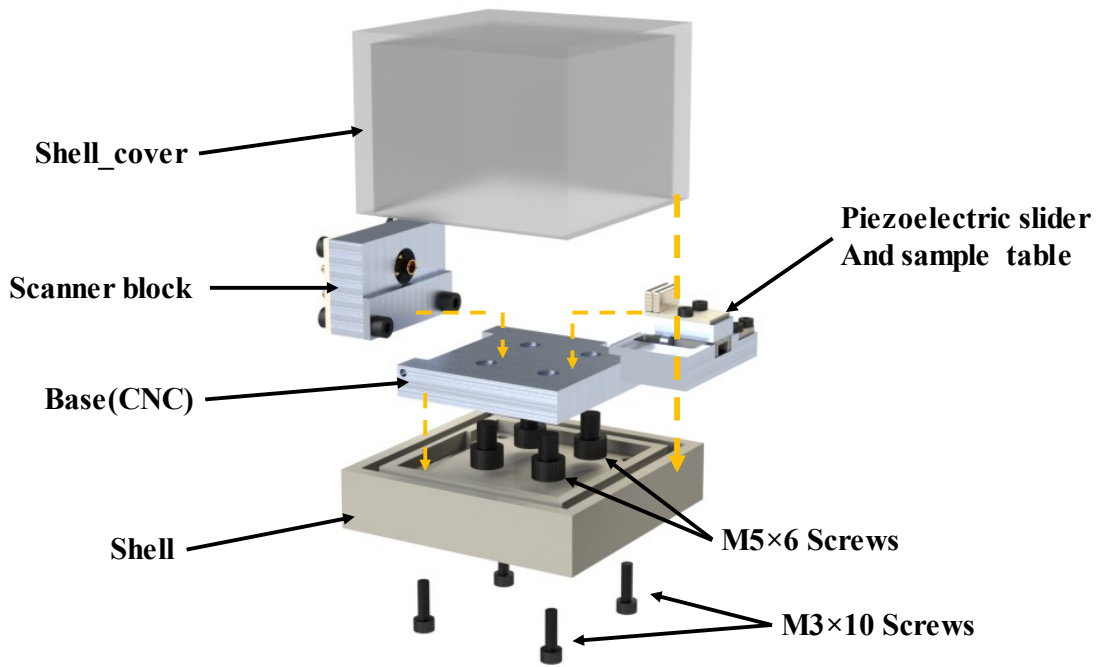

Fig. 11. Schematic of the leading microscope assembly. The **shell** and **shell_cover** together form the dust cover.

## 5.4 Wiring

Please refer to Table. 2 to establish connections between PCBs using different types of coaxial cables. The coaxial cables should pass through the wire groove on the shell when connecting the control unit to the STM body.

Table. 2. Wiring table

| Source | Target | Source Silkscreen | Target Side Silkscreen | Function Describe | Wire type |
|---|---|---|---|---|---|
| Power Board | Control Board | V_LDO_-12VA | -12V_IN | DAC Power | SMA-SMA |
| Power Board | Control Board | V_LDO_-12VB | +12V_IN |  | SMA-SMA |
| Power Board | Scanner_Connector | V_LDO_+12VA | +12V | Pre-amplifier power | SMA- IPEX |
| Power Board | Scanner_Connector | V_LDO_+12VB | -12V |  | SMA- IPEX |
| Power Board | Control Board | V_LDO_+5V | +5V_IN | ADC Power | SMA-SMA |
| MCU Board | Power Board | PWR_OUT | BOOSTER_IN | 5V Power | SMA-SMA |
| Control Board | Scanner_Connector | Z-Y | Z-Y | Scanner control | SMA- IPEX |
| Control Board | Scanner_Connector | Z+Y | Z+Y |  | SMA- IPEX |
| Control Board | Scanner_Connector | Z-X | Z-X |  | SMA- IPEX |
| Control Board | Scanner_Connector | Z+X | Z+X |  | SMA- IPEX |
| Control Board | Sample_ConnectBoard | 16Bit-DAC | BIAS | Sample bias voltage | SMA- IPEX |
| Control Board | Sample_ConnectBoard | 12Bit-DAC | PZS | Piezoelectric slider driver | SMA- IPEX |
| Control Board | Scanner_Connector | ADC_IN | OUTPUT | Pre-amplifier output | SMA- IPEX |

## 5.5 Tip and sample preparation

In an ideal scenario, an STM tip typically consists of only a few atoms. The standard procedure for tip preparation involves electrochemical etching using a tungsten wire[20]. This method can yield a tip with significant sharpness, but the preparation process is time-consuming and complex. In practice, using platinum-iridium alloy wire and cutting it can also produce tips that meet the requirements of an STM[21]. Platinum is an inert metal that can effectively prevent oxidation after cutting, and the alloy of platinum and iridium further enhances the tip's hardness. To make the tip easier to manufacture, we attempted to use pure platinum wire for tip preparation, and the results were successful. The fabrication method involves wiping a 0.3mm diameter platinum wire with alcohol, securing one end of the platinum wire with tweezers, and stretching and cutting the other end at an angle of about 45 degrees using wire cutters. The cut tip is then inserted into the MMCX connector of the scanner.

The sample is made using a magnet(square). The steps are as follows: First, glue the test material onto the magnet, then apply conductive silver paste between the sample and the magnet to establish an electrical connection. After the conductive silver paste has dried, attach the prepared sample to the sample table using magnetic adsorption, and then the scanning can proceed.

## 6 Operation instructions

This section will describe the preparations before microscope imaging, software operations, and adjustments to parameters for the piezo stage and approach algorithm.

### 6.1 Hardware setting up

(1) Use jumper caps to connect the **Control Board** jumpers to power the DACs and ADCs: H1, H2, and H3.

(2) Use jumper caps to connect all jumpers named PWR on the **Power Board**, connect the SS and V_EN jumper, connect COMP1 and COMP2 to the GND terminal, SYNC/FREQ to the 2.4M terminal, SEQ to the SEQ. Terminal, SLEW to the NORM terminal, EN1 to the ON terminal, and EN2 to the OFF terminal.

(3) Connect the jumpers PWR_DEBUG_EN and USB_EN on the **MCU Board**.

(4) Connect a low-ripple USB 5V Type-C power supply to the USB-C PWR port. Connect the USB on the computer to the USB_DEBUG port.

(5) Use tweezers to insert the tip into the scanner.

(6) Load the sample.

(7) Move the piezoelectric slider to bring the sample close to the tip, about 1mm.

(8) Cover it with the dust cover and complete the scanning on the software side.

### 6.2 Software interface

The software interface comprises five sections: UART Connect box, Slider Approach Control box, Tunneling Current Monitoring box, and a multi-page testing tab. A typical testing process includes the following steps:

(1) Click Refresh in the UART Connect box to select the corresponding serial port.

(2) Click the approach button in the Approach box and wait for the approach to complete.

(3) In the multi-page Testing Tab, select the appropriate testing category, adjust parameters, and complete the test.

(4) Save the corresponding file(Curve data or image).

Curve data is saved as an Excel file, while images are saved as TIFF files. Image processing can be performed using other software, such as Gwiddion.

### 6.3 Piezoelectric slider setting

Liao et al. described the positioner's operating principle based on the stick-slip phenomenon[8]. When the positioner operates, it can be divided into two modes: high-resolution mode (applying a triangular wave to the piezo stack) and low-resolution mode (using a sawtooth wave to the piezo stack). Since we don't need a high resolution in the pre-approach phase and we need to save time, our slider only operates in the low-resolution mode.

Fig. 12 illustrates the working principle of the piezoelectric slider. The piezo stack elongates when the piezo voltage rises, causing the slider to move to the right by ΔS1 due to the magnet's influence. However, the piezo stack rapidly contracts when the piezo voltage returns to zero. Due to the stick-slip phenomenon[22], the slider cannot wholly follow the movement of the piezo stack, resulting in a displacement of ΔS3. In this process, ΔS1 represents the displacement when the slider moves synchronously with the piezo stack, ΔS2 is the displacement generated by the contraction of the piezo stack driving the slider, and ΔS1 = ΔS2 + ΔS3.

In the software settings page, the Slider Amplitude parameter can be modified to control ΔS1 of the slider. A larger ΔS1 will result in a faster speed but may increase the risk of the tip crashing.

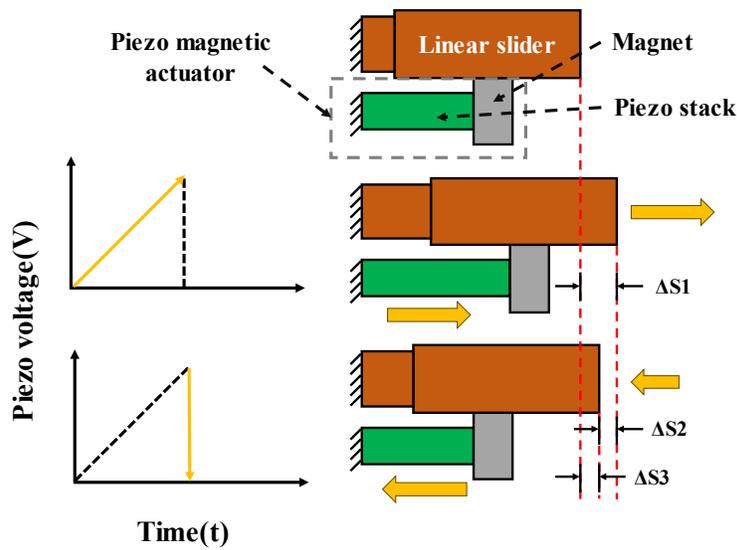

Fig. 12. Piezoelectric slider working diagram in low-resolution mode

## 6.4 Approach method

We observed that during the movement of the piezoelectric slider, there is a fluctuation in the output signal of the pre-amplifier when the tip gets sufficiently close to the sample. Fig. 13(a) clearly illustrates this fluctuation. The blue dashed line represents the piezo stack driving voltage, where negative voltage pushes the sample toward the tip. The pre-amplifier detects a positive current in the gray region as the distance between the tip and the sample shortens. In the green area, during the phase of relative sliding between the piezo stack and the slider, the current suddenly reverses and exhibits oscillation. To further determine the source of this current, we let the slider continue moving and captured the curve change just before the tip collision, as shown in Fig. 13(b). By taking the logarithm of the curve, we found that the current exhibits exponential changes, suggesting that the tunneling current dominates the gray region's current.

Since the oscillation region undergoes polarity reversal, it rules out the possibilities of breakdown, electron emission, and tunneling principles [23]. Considering that the capacitance between the tip and the sample can generate a detectable signal [15][24], we speculate that the oscillation in this part is due to the brake moan oscillation generated by the friction between the magnet on the piezo stack and the slider [25]. In fact, in Fig. 13(b), this oscillation can also be observed, and its frequency is close to that in Fig. 13(a). As the distance between the sample and the tip decreases, this oscillation becomes more pronounced. We use this signal as a characteristic value to estimate the sample-tip distance and enhance the speed of the pre-approach procedure. Fig. 13(c) shows the current change during continuous slider movement, and it can be seen that as the slider moves, the oscillation amplitude becomes more significant.

The entire approach is divided into three stages: fast mode, slow mode, and Z scan mode. In fast mode, the slider moves at a set speed and detects the peak of current oscillation after each step. It enters slow mode if it reaches a set threshold (Fast Cap Current). In slow mode, the slider's movement speed slows down, and it enters the Z Scan mode when the oscillation peak reaches the second threshold. In Z Scan Mode, with each step, the system controls the scanner in the Z-direction to determine if the sample has entered the scanner's scanning range. The approach is completed if the tunneling current reaches the set point during this process. The threshold parameters mentioned above can all be configured in the settings tab. With the parameters correctly set, loading the sample takes approximately 1 minute to establish the tunneling current.

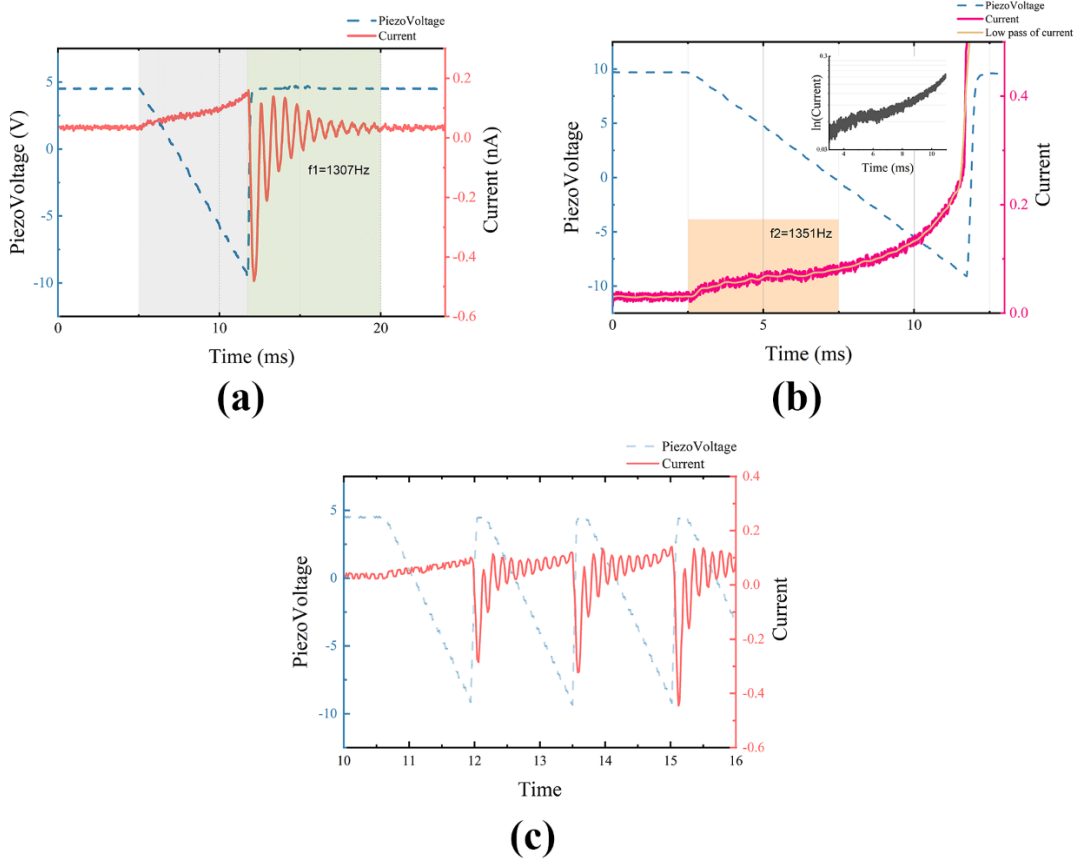

Fig. 13. Pre-amplifier (TIA) output signal vs piezoelectric slider voltage, sample bias voltage $|V_{bias}|=10V$, Highly Oriented Pyrolytic Graphite (HOPG) sample from JH Special Carbon Technology Co., LTD, the curve is sampled using a RIGOL 1102Z-E oscilloscope. (a) Response of the TIA to a single-step movement of the slider. (b) Response of the TIA to a single-step slider movement, with tip collision occurring after this step. (c) Response of the TIA to continuous multiple-step movements of the tip.

# 7   Validation and characterization

We used a 0.5mm thick Highly Oriented Pyrolytic Graphite (HOPG) sample from JH Special Carbon Technology Co., LTD for sample preparation and conducted tests on it. The test results presented below were obtained using the described vibration isolation system in an office environment on the 4th floor.

Fig. 14 shows our test results for the bias curve of HOPG. After the sample and tip reached the tunneling distance, we varied the sample bias voltage from -2V to 2V and recorded the current output. Compared to the curves obtained by other researchers[26][27], the characteristics of the curve we measured are similar to theirs.

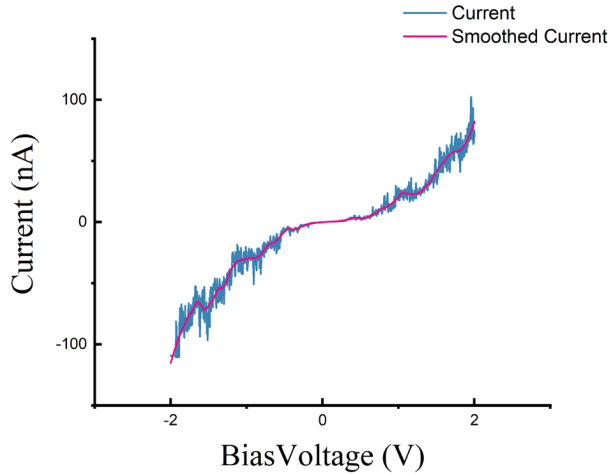

Fig. 14. Bias voltage vs tunneling current of HOPG, range from -2V to 2V

Next, we performed STM imaging of the HOPG sample. Fig. 15(a) shows the raw image of HOPG with a scan size of 50×50, a sample bias voltage of 80mV, and a scan time of 657ms. From Fig. 15(a), we can observe the triangular structure of graphite atoms in the STM image pattern. Fig. 15(b) represents the result after performing a two-dimensional fast Fourier transform on the raw image. It is dominated by six Fourier coefficients with an angle of approximately 120°, consistent with the behavior of HOPG under scanning tunneling microscopy [28].

By varying the bias voltage, the number of sampling points, and swapping the tip, we conducted multiple scans of the same HOPG sample, all of which revealed the atomic characteristics of HOPG. Fig. 16(a) shows the result of a sample bias voltage of 50mV and a scan size of 60×60, displaying a more pronounced lattice structure.

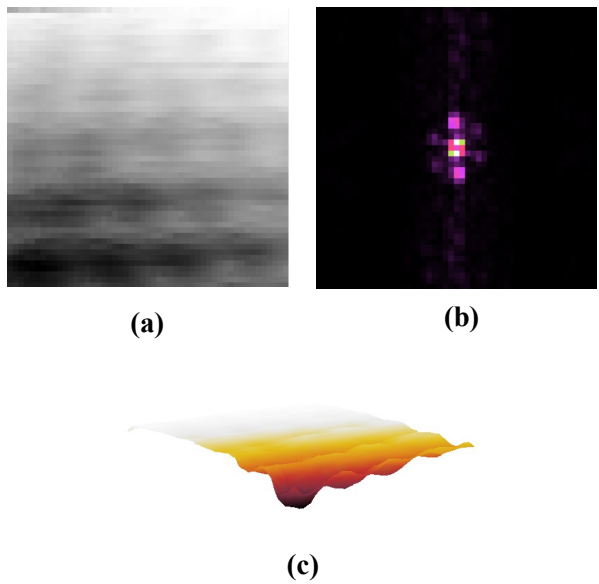

Fig. 15 Raw scanning tunneling microscope image of HOPG, 50×50 sample area, X/Y scan voltage range from 0~152.5mV,|$V_{bias}$|=80mV, scan time=657ms. (a) 2D gray level image (b) FFT of raw image (c) 3D view of the image

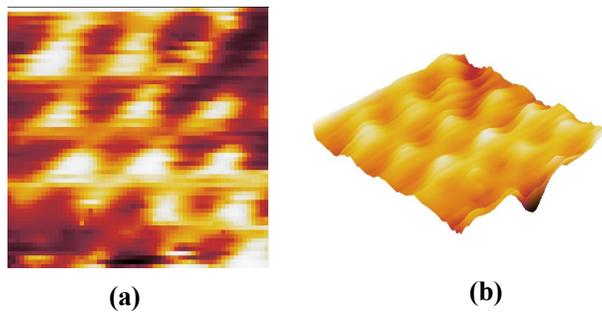

Fig. 16 Scanning tunneling microscope image of HOPG, 60×60 sample area, X/Y scan voltage range from 0~183mV,|Vbias|=50mV. (a) Proceed HOPG image with Gwiddion (b) 3D view of HOPG image

## Acknowledgments

Thanks to Mr. Jürgen Müller from Germany for his advice on this project. We also appreciate the support provided by Shenzhen JLC Technology Group Co.,Ltd. for this project, as well as all the suggestions from the Bilibili audience.

## References

[1]  Binnig, G., & Rohrer, H. (2000). Scanning tunneling microscopy. *IBM Journal of research and development*, 44(1/2), 279.

[2] Binnig, G., & Rohrer, H. (1987). Scanning tunneling microscopy—from birth to adolescence. *Reviews of modern physics*, 59(3), 615.

[3] The Scanning Tunneling Microscope (STM) - The Advantages of Atomic Resolution. (n.d.). *MicroscopeMaster*. Retrieved September 15, 2023, from https://www.microscopemaster.com/scanning-tunneling-microscope.html

[4] Besocke, K. (1987). An easily operable scanning tunneling microscope. *Surface Science*, 181(1-2), 145-153.

[5] Alexander, J. D. (n.d.). *Simple STM Design Page*. Simple STM Design Page. Retrieved September 15, 2023, from https://john-alexander42.github.io/simple-stm-web-page/index.htm

[6] Alexander, J. D., Tortonese, M., & Nguyen, T. (1999). *U.S. Patent No. 5,952,657*. Washington, DC: U.S. Patent and Trademark Office.

[7] Berard, D. (2015, January 12). *A home-built scanning tunneling microscope with atomic resolution. A Home-built Scanning Tunneling Microscope With Atomic Resolution* | Dan Berard. Retrieved September 15, 2023, from https://dberard.com/2015/01/12/a-home-built-scanning-tunneling-microscope-with-atomic-resolution/

[8] Liao, H. S., Werner, C., Slipets, R., Larsen, P. E., Hwang, S., Chang, T. J., ... & Hwu, E. T. (2022). Low-cost, open-source XYZ nanopositioner for high-precision analytical applications. *HardwareX*, 11, e00317.

[9] Chen, C. J. (2021). Introduction to Scanning Tunneling Microscopy Third Edition (Vol. 69). *Oxford University Press*, USA.

[10] Grafstrom, S., Kowalski, J., & Neumann, R. (1990). Design and detailed analysis of a scanning tunnelling microscope. *Measurement Science and Technology*, 1(2), 139.

[11] Park, S. I., & Quate, C. F. (1987). Theories of the feedback and vibration isolation systems for the scanning tunneling microscope. *Review of scientific instruments*, 58(11), 2004-2009.

[12] Okano, M., Kajimura, K., Wakiyama, S., Sakai, F., Mizutani, W., & Ono, M. (1987). Vibration isolation for scanning tunneling microscopy. *Journal of Vacuum Science & Technology A: Vacuum, Surfaces, and Films*, 5(6), 3313-3320.

[13] Binnig, G., Rohrer, H., Gerber, C., & Weibel, E. (1982). Tunneling through a controllable vacuum gap. *Applied Physics Letters*, 40(2), 178-180.

[14] Petersen, J. P., & Kandel, S. A. (2017). Circuit design considerations for current preamplifiers for scanning tunneling microscopy. *Journal of Vacuum Science & Technology B*, 35(3).

[15] De Voogd, J. M., Van Spronsen, M. A., Kalff, F. E., Bryant, B., Ostojić, O., Den Haan, A. M. J., ... & Rost, M. J. (2017). Fast and reliable pre-approach for scanning probe microscopes based on tip-sample capacitance. *Ultramicroscopy*, 181, 61-69.

[16] Electronics. (2014, December 29). *Electronics | Dan Berard*. Retrieved September 15, 2023, from https://dberard.com/home-built-stm/electronics/

[17] Stipe, B. C., Rezaei, M. A., & Ho, W. (1999). A variable-temperature scanning tunneling microscope capable of single-molecule vibrational spectroscopy. *Review of Scientific Instruments*, 70(1), 137-143.

[18] Wong, T. M. H., & Welland, M. E. (1993). A digital control system for scanning tunnelling microscopy and atomic force microscopy. *Measurement Science and Technology*, 4(3), 270.

[19] OPA627 data sheet, product information and support | TI.com. (n.d.). *OPA627 Data Sheet, Product Information*

*and Support | TI.com*. Retrieved September 15, 2023, from https://www.ti.com/product/OPA627

[20] Melmed, A. J. (1991). The art and science and other aspects of making sharp tips. *Journal of Vacuum Science & Technology B: Microelectronics and Nanometer Structures Processing, Measurement, and Phenomena*, 9(2), 601-608.

[21] Gorbunov, A. A., Wolf, B., & Edelmann, J. (1993). The use of silver tips in scanning tunneling microscopy. *Review of scientific instruments*, 64(8), 2393-2394.

[22] Gao, C., Kuhlmann-Wilsdorf, D., & Makel, D. D. (1994). The dynamic analysis of stick-slip motion. *Wear*, 173(1-2), 1-12.

[23] Wallash, A. J., & Levit, L. (2003, January). Electrical breakdown and ESD phenomena for devices with nanometer-to-micron gaps. In *Reliability, testing, and characterization of MEMS/MOEMS II* (Vol. 4980, pp. 87-96). SPIE.

[24] Schlegel, R., Hänke, T., Baumann, D., Kaiser, M., Nag, P. K., Voigtländer, R., ... & Hess, C. (2014). Design and properties of a cryogenic dip-stick scanning tunneling microscope with capacitive coarse approach control. *Review of Scientific Instruments*, 85(1).

[25] Nack, W. V., & Joshi, A. M. (1995). Friction induced vibration: brake moan. SAE transactions, 1967-1973.

[26] El Abedin, S. Z., Borissenko, N., & Endres, F. (2004). Electrodeposition of nanoscale silicon in a room temperature ionic liquid. *Electrochemistry communications*, 6(5), 510-514.

[27] Wang, Q., Hou, Y., Wang, J., & Lu, Q. (2013). A high-stability scanning tunneling microscope is achieved by an isolated tiny scanner with low voltage imaging capability. *Review of Scientific Instruments*, 84(11).

[28] Mizes, H. A., Park, S. I., & Harrison, W. A. (1987). Multiple-tip interpretation of anomalous scanning-tunneling-microscopy images of layered materials. *Physical Review B*, 36(8), 4491.